\documentclass[11pt,twoside]{article}

\usepackage{asp2006}
\usepackage{epsf}
\usepackage{epsfig}
\usepackage{lscape}
\usepackage{graphicx}
\usepackage{amssymb, amsmath, amsbsy}
\newcommand{\lbol}{\ensuremath{L\mathrm{_{bol}}}}

\newcommand{\ledd}{\ensuremath{L\mathrm{_{Edd}}}}
\newcommand{\lratio}{\ensuremath{L/\ledd}}

\newcommand{\msun}{\ensuremath{M_{\odot}}}

\newcommand{\mbh}{\ensuremath{M_\mathrm{BH}}}

\newcommand{\st}{\ensuremath{\sigma_{\rm \scriptscriptstyle T}}}
\newcommand{\massp}{\ensuremath{m_{\rm p}} }

\newcommand{\rs}{\ensuremath{r_{\rm \scriptscriptstyle S}}}

\newcommand{\hb}{H\ensuremath{\beta}}

\newcommand{\oiii}{[O\,III]}

\newcommand{\feii}{Fe\,II}

\newcommand{\mgii}{Mg\,II}

\newcommand{\civ}{C\,IV}
\newcommand{\lya}{Ly$\alpha$}

\newcommand{\colnh}{\ensuremath{N_\mathrm{H}}}

\begin{document}

\title{The origin and physical mechanism of the ensemble Baldwin effect} 
\author{Xiaobo Dong\altaffilmark{1}, Jianguo Wang\altaffilmark{2,1},
Tinggui Wang\altaffilmark{1}, Huiyuan Wang\altaffilmark{1},
Xiaohui Fan\altaffilmark{3}, Hongyan Zhou\altaffilmark{1},
Weimin Yuan\altaffilmark{2}, and Qian Long\altaffilmark{4}  }

\affil{}    
\altaffiltext{1}{Center for Astrophysics, University of Science and
Technology of China, Hefei, Anhui 230026, China; xbdong@ustc.edu.cn}
\altaffiltext{2}{National Astronomical Observatories/Yunnan
Observatory, Chinese Academy of Sciences, P.O. Box 110, Kunming,
Yunnan 650011, China}
\altaffiltext{3}{Steward Observatory, The University of Arizona,
Tucson, AZ 85721}
\altaffiltext{4}{Department of Precision Machinery and Instrumentation,
University of Science and
Technology of China, Hefei, Anhui 230026, China}

\begin{abstract} 
We have conducted a systematic investigation of the origin and underlying physics
of the line--line and line--continuum correlations of AGNs, particularly the
Baldwin effect.
Based on the homogeneous sample of Seyfert 1s and QSOs in the SDSS DR4,
we find the origin of all the emission-line regularities is Eddington ratio (\lratio).
The essential physics is that \lratio\ regulates the distributions of
the properties (particularly column density) of the clouds
bound in the line-emitting region.
\end{abstract}


\section{Baldwin effect and its three 2nd-order effects}

Active galactic nuclei (AGNs; including QSOs)
are essentially a kind of radiation and line-emitting systems
powered by gravitational accretion onto suppermassive black holes.
The global spectra of AGNs are remarkably similar,
emission lines with similar intensity ratios sitting atop a blue continuum
that has a similar slope among AGNs regardless of their luminosities and redshifts
(Davidson \& Netzer 1979, Korista 1999).
However, this similarity is only a zeroth-order approximation;
in 1977, Baldwin found that, among the QSO ensemble,
the equivalent width (EW) of the \civ\ $\lambda$1549 emission line
correlates negatively with the continuum luminosity (Baldwin 1977).
From then on, such a negative EW--L correlation has been found for almost
all emission lines in the ultraviolet and optical bands, and has been termed
as ``Baldwin effect'' (hereafter BEff; see Osmer \& Shields 1999, Sheilds 2007
for reviews).\footnote{To distinguish the BEff in the ensemble from a similar
correlation present in individual variable AGNs, the former is usually called
``the ensemble BEff'' that is the topic of this paper.}
Furthermore, also found are the three kinds of 2nd-order effects of the BEff,
namely, the dependence of the BEff slope
(the slope of the log\,EW--log\,L relation)
on luminosity, 
on ionization energy, 
and, particularly, on velocity.
The \civ\ BEff is stronger in the peak and red wing than in the blue
(Francis \& Koratkar 1995, Richards et al. 2002)!
The velocity dependence betrays the nature of the BEff:
this kind of negative correlation relates to the line-emitting gas gravitationally bound in
the broad-line region (BLR).

\section{The origin and the physical mechanism}
In order to explore the origin and the underlying mechanism of the BEff,
based on the homogeneous sample of 4178 $z\leq 0.8$ Seyfert 1s and QSOs
with median spectral S/N $\gtrsim 10$ per pixel in the SDSS DR4,
we have conducted a systematic investigation of the line--line and line--continuum
correlations for broad and narrow emission lines in the near-UV and optical,
from \mgii\ $\lambda 2800$ to \oiii\ $\lambda 5007$.
Our findings are as follows:\\
\indent (i) The strongest correlations of almost all
the emission-line intensity ratios and EWs are with \lratio,
either positively (e.g. \feii\ EW) or negatively (e.g. \mgii\ EW),
rather than with $L$ or \mbh;
besides, generally intensity ratios have tighter correlations with \lratio\ than
the EWs of the related lines. \\
\indent (ii) The intensity ratios of \feii\ emissions -- both narrow and broad --
to \mgii\ have very strong, positive correlations with \lratio;
interestingly enough, (narrow \feii\ $\lambda 4570$)/\mgii\ has a stronger correlation
with \lratio\ than the optical and UV (broad \feii)/\mgii, with Spearman \rs = 0.74
versus 0.58 (optical) and 0.46 (UV); see Fig. 1 (Dong et al. 2009a).

These findings argue that Eddington ratio ($\ell \equiv \lratio$)%
\footnote{
Eddington ratio is the ratio between the bolometric and
Eddington luminosities.
Eddington luminosity (\ledd), by definition, is the luminosity at which the gravity
of the central source acting on an electron--proton pair (i.e. fully ionized gas)
is balanced by the radiation pressure due to electron Thomson scattering;
$\ledd = 4 \pi G c M \massp/ \st $,
where $G$, $c$, $M$, \massp, \st\
are the gravitational constant, speed of light, mass of the central source,
proton mass, Thomson scattering cross-section, respectively.
In accretion-powered radiation systems, \lratio\ is often
referred to as \emph{dimensionless accretion rate $\dot{m}$}
(the relative accretion rate normalized by Eddington accretion rate $\dot{M}_{\rm Edd}$,
$\dot{m} \equiv \dot{M}/\dot{M}_{Edd} = \eta c^2 \dot{M}/ \ledd$,
$\dot{M}$ being mass accretion rate and $\eta$ the accretion efficiency)
as $\dot{m}$ is not an observable; yet the two notations are different
both in meaning and in scope of application.
Even in the accretion-powered radiation systems like AGNs,
\lratio\ ($L$) is not equivalent to $\dot{m}$ ($\dot{M}$)
except in the simple thin accretion disk model of Shakura \& Sunyaev (1973).
Therefore, we would rather call \lratio\
\emph{dimensionless luminosity} ($\ell$).
}
is the origin of the BEff,
as of other regularities of almost all emission lines
(e.g., the \feii--\oiii\ anticorrelation, Boroson \& Green 1992).
This once has been suggested by
Baskin \& Laor (2004) and Bachev et al. (2004)
for the \civ\ BEff.
We propose that the underlying physics is certain
self-regulation mechanisms
caused by (or corresponding to) \lratio;
these mechanisms maintain the normal dynamically quasi-steady
states of the gas surrounding
the central engine of AGNs (Dong et al. 2009a,b).
Briefly, the essential one is that
\emph{there is a lower limit on the column density (\colnh) of the clouds
gravitationally bound in the AGN line-emitting region, set by \lratio}
(hereafter the \colnh--\lratio\ mechanism;
see also Fig. 1 of Fabian et al. 2006, Marconi et al. 2008).
As \lratio\ increases, the emission strength decreases for
high-ionization lines (e.g. \civ) and
optically thick lines
that are emitted at the illuminated surface (e.g. \lya)
or in the thin transition layer (e.g. \mgii)
of the BLR clouds;
for low-ionization, optically thin lines such as \feii\ multiplets
that originate from the volume behind the Hydrogen ionization front
(i.e., from the ionization-bounded clouds only),
as \lratio\ increases the emission strength increases.%
\footnote{
Certainly, for particular radiation systems that are powered
by gravitational accretion (e.g. AGNs),
$\ell$ is linked tightly to $\dot{m}$ anyway
(cf. Footnote 2; Merloni \& Heinz 2008).
Thus there is an additional effect associated with $\ell$ yet directly via $\dot{m}$
as follows.
The increase in $\ell$ means, meanwhile,
the increase in $\dot{m}$, the gas supply.
Reasonably, it is from the supplied gas spiraling into the central engine
that (at least a significant part of) the line-emitting clouds originate;
this is particularly true for the BLR and inner NLR clouds
that are located between the torus (as the fuel reservoir) and the accretion disk
(e.g. Gaskell \& Goosman 2008).
Hence, as $\ell$ increases, the total mass
of line-emitting gas increases.}
This is schematically sketched in Fig. 2.

\section{The implications}
~~~~~~$\mathcal{I.}$ An implication is that BG92's PC1, if only the spectral correlations
in the UV--optical are concerned,
shares the same origin with PC2 that is exactly the He\,II BEff.
A lesson is that we should be more cautious about the premises of
blind source separation methods such as Principal Component Analysis.

$\mathcal{II.}$ As suggested insightfully by G.~Richards (e.g. Richards 2006),
\emph{the \civ\ line blueshifting (in other words, blue asymmetry)
is the same phenomenon of BEff}.
The underlying physical picture is clear now:
There are two components in the \civ\ emission, one arising from outflows
and the other from the clouds gravitationally bound in the BLR;
the fraction of bound clouds that optimally emit \civ\ line decreases
with increasing \lratio\ according to the \colnh\ -- \lratio\ mechanism.

$\mathcal{III.}$ If the observed large scatter of \feii/\mgii\ at the same redshift
is caused predominately by the diversity of \lratio,
then once this systematic variation is corrected according to the tight
\feii/\mgii\ -- \lratio\ correlation,
it is hopeful to still use \feii/\mgii\ as a measure of the Fe/Mg abundance ratio
and thus a cosmic clock (\emph{at least in a statistical manner}).

\appendix
\section*{Appendix: Not \emph{Baldwin Effect}, but \emph{ell Effect}?}
This Appendix is to present more results taken from Dong et al. (2009b) that did not
appear in the proceedings paper due to page limit.
The aim is to show that the traditional BEff -- the dependence of emission-line EWs
on luminosity -- is likely not to be fundamental, but derived from the dependence on
Eddington ratio ($\ell$).
To avoid any confusion, below we call the latter
ell effect (\emph{Ell} stands for $\ell$, Eddingtion ratio).

The ell effect is the 1st-order small variation to the 0th-order
global similarity of QSO spectra that is well explained by ``locally
optically-emitting clouds'' (LOC) photoionization modeling (Baldwin et al. 1995).
As discussed above, in the study of the QSO emission-line correlations,
the focus seems to be shifting from the physics (\emph{microphysics})
mainly of the accretion process
to the `statistical physics' (\emph{macrophysics})
of the surrounding clouds (Korista 1999; Korista, private communication).
With more realistic constraints to be accounted for [e.g., the
distribution function of the cloud number with $\ell$ and $\colnh$,
$N_{\rm c}(\ell, \colnh(\ell)\,)$; cf. Fig. 2], the 1st-order
regularities and even the 2nd-order effects of QSO emission lines
(see \S1) may be reproduced exactly by future LOC modeling.

Fig. 3 confirms that $\ell$ is the primary driver of the BEff of \mgii\ $\lambda 2800$.
Fig. 4 shows that, at the 0th-order approximation,
\mgii\ luminosity is directly proportional to the continuum luminosity, exactly as
predicted by photoionization theory;
at the 1th-order approximation, for different $\ell$
the proportional coefficient is different,
$ \log k \simeq \log k_0 + k' \cdot \log \ell $ ---
this is just the ell effect illustrated in Fig. 3c.

Some researchers once have found that
the slopes of the emission line versus continuum luminosity relations
in the log--log scale is not unity (see references in \S2 of Shields 2007).
We must note that this is likely not to be intrinsic (see Dong et al. 2009b for
a detailed investigation).
It is caused partly by selection effect inherent in any magnitude-limited sample,
with high-luminosity objects having higher $\ell$ and thus smaller EWs.
For optical emission lines particularly (e.g. \hb), this is mainly caused by
the contamination of the host-galaxy starlight (cf. Croom et al. 2002).
The starlight contamination aggravates gradually towards longer wavelengths;
moreover, within a fixed aperture, it aggravates with decreasing AGN luminosity.

In one word, a sole fundamental parameter, $\ell$,
well regulates the ordinary state of the surrounding gas
that is inevitably inhomogeneous and clumpy (`clouds').

\begin{figure}[tbp]
\centering
\includegraphics[width=1\textwidth]{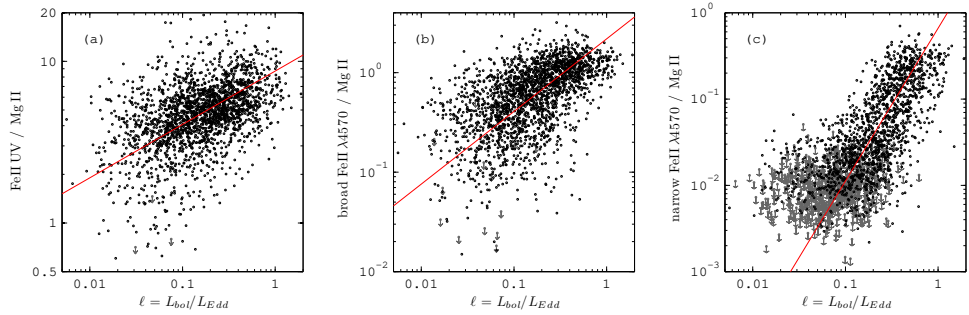}
\caption{
Plots of the intensity ratios of the broad and narrow \feii\
emissions to \mgii\ $\lambda 2800$ versus Eddington ratio
for the homogeneous sample of 4178 Seyfert 1s and QSOs.
Also plotted are the best-fitted linear relations in the log--log scale
(Dong et al. 2009a). }
\end{figure}

\begin{figure}
\centering
\includegraphics[width=1\textwidth]{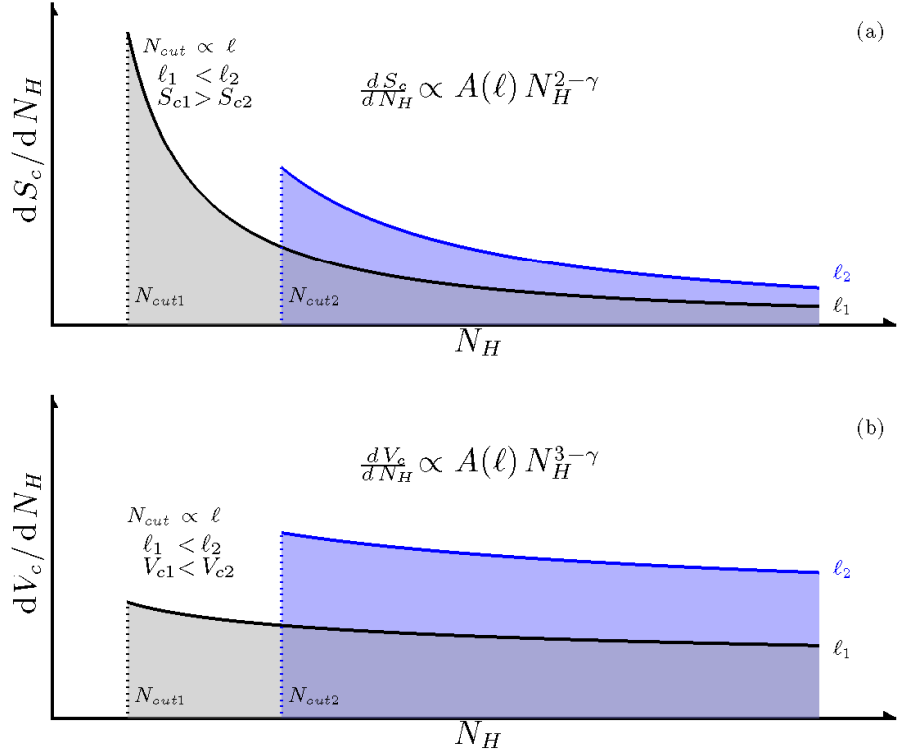}
\caption{Schematic sketches of the distribution functions of the
total illuminated surface area $S_\mathrm{c}$ with \colnh\ (panel a)
and of the total volume $V_\mathrm{c}$ with \colnh\ (panel b) of the
clouds bound in the broad-line region and inner narrow-line region
of AGNs. The sketches are plotted assuming the distribution of the
cloud number to be $\mathrm{d}\,N_\mathrm{c}\,/\,\mathrm{d}\,\colnh
= A(\ell)\, {\colnh}^{-\gamma}$ with $\gamma = 3.2$;
the lower-limit \colnh\ cutoff is set by the \colnh--\lratio\ mechanism, roughly
scaling with \lratio.
Note that the strengths of high-ionization lines (e.g. \civ) and
optically-thick lines (e.g. \lya\ and \mgii) are roughly
proportional to $S_\mathrm{c}$ while that of low-ionization, optically-thin lines
such as narrow-line and broad-line \feii\
roughly proportional to $V_\mathrm{c}$ (Dong et al. 2009b).
}
\end{figure}

\begin{figure}[tbp]
\centering
\includegraphics[width=1\textwidth]{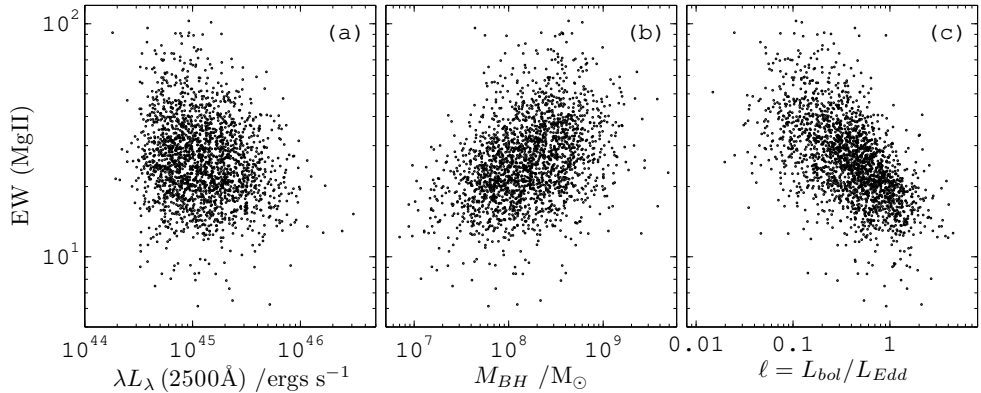}
\caption{
The relation between the equivalent width of \mgii\ $\lambda 2800$
of the 2092 type-1 AGNs in the $z>0.45$ subsample
and their $\lambda L_{\lambda}$(2500\AA), black hole mass (\mbh),
and Eddington ratio ($\ell \equiv \lbol/\ledd$).}
\end{figure}

\begin{figure}[tbp]
\centering
\includegraphics[width=1\textwidth]{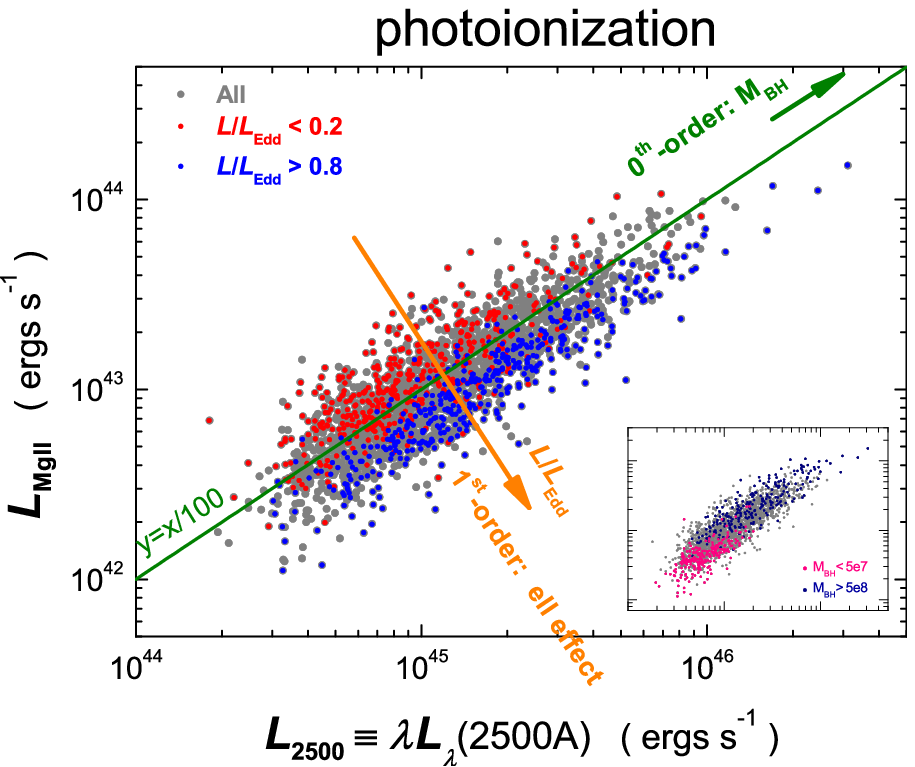}
\caption{
The log--log plot of \mgii\ versus continuum luminosity relation
for the 2092 objects.
The green line is the 1-to-100 relation to aid the eye.
Inset is the plot of the same data,
with objects having $\mbh < 5 \times 10^7$ \msun\ denoted as pink and
objects having $\mbh > 5 \times 10^8$ \msun\ navy-blue.
Note that $L_{\rm MgII} = k(\ell) \, L_{2500}
\simeq k_0 L_{2500} \, \ell ^{k'} $.
The 0th-order term, $k_0$, can be calculated by
the classical LOC photoionization of Baldwin et al. (1995);
the 1st-order term, $\ell ^{k'}$, by the ell effect (cf. Fig. 3c).
See Dong et al. (2009b) for details. }
\end{figure}



\acknowledgments
DXB thanks Kirk Korista, Martin Gaskell, Aaron Barth, Alessandro Marconi,
Philippe V\'{e}ron,
Daniel Proga and Xueguang Zhang for the helpful discussions and comments,
thanks Zhen-Ya Zheng for the help in improving IDL figures,
and thanks Sheng-Miao Wu, Lei Chen and Fu-Guo Xie for their warm hospitality
and brainstorming discussions during my visits in Shanghai Observatory.
This work has made use of the data of
the Sloan Digital Sky Survey (SDSS).
The SDSS Web Site is http://www.sdss.org/.
This work is supported by Chinese NSF grants
NSF-10533050, NSF-10703006 and NSF-10728307,
the CAS Knowledge Innovation Program (Grant No. KJCX2-YW-T05),
and a National 973 Project of China (2007CB815403).





\begin{thebibliography}{}

\bibitem[Bachev et al.(2004)]{2004ApJ...617..171B} Bachev, R., et al.\ 2004, \apj, 617, 171

\bibitem[Baldwin(1977)]{1977ApJ...214..679B} Baldwin, J.~A.\ 1977, \apj,
214, 679

\bibitem[Baldwin et al.(1995)]{1995ApJ...455L.119B} Baldwin, J., Ferland,
G., Korista, K., \& Verner, D.\ 1995, \apjl, 455, L119

\bibitem[Baskin \& Laor(2004)]{2004MNRAS.350L..31B} Baskin, A., \& Laor, A.\ 2004, \mnras, 350, L31

\bibitem[Boroson \& Green(1992)]{1992ApJS...80..109B} Boroson, T.~A.~\&
Green, R.~F.\ 1992, \apjs, 80, 109 (BG92)


\bibitem[Croom et al.(2002)]{2002MNRAS.337..275C} Croom, S.~M., et al.\
2002, \mnras, 337, 275



\bibitem[Davidson \& Netzer(1979)]{1979RvMP...51..715D} Davidson, K., \&
Netzer, H.\ 1979, Reviews of Modern Physics, 51, 715

\bibitem[Dong et al.(2009a)]{} Dong, X., Wang, J., Wang, T., Wang, H., Fan, X.,
Zhou, H., Yuan, W.\ 2009a,
arXiv:0903.5020

\bibitem[Dong et al.(2009b)]{} Dong, X., Wang, J., Wang, T., Wang, H., Fan, X.,
Zhou, H., Yuan, W., Qian, L.\ 2009b, \apj\ to be submitted

\bibitem[Fabian et al.(2006)]{2006MNRAS.373L..16F} Fabian, A.~C., Celotti,
A., \& Erlund, M.~C.\ 2006, \mnras, 373, L16



\bibitem[Francis
\& Koratkar(1995)]{1995MNRAS.274..504F} Francis, P.~J., \& Koratkar, A.\ 1995, \mnras, 274, 504




\bibitem[Gaskell \& Goosmann(2008)]{blr-torus}
Gaskell, C.~M., \& Goosmann, R.~W.\ 2008, arXiv:0805.4258


\bibitem[Korista(1999)]{1999ASPC..162..429K} Korista, K.\ 1999, Quasars and
Cosmology, ASPC, 162, 429


\bibitem[Marconi et al.(2008)]{2008ApJ...678..693M} Marconi, A., Axon,
D.~J., Maiolino, R., Nagao, T., Pastorini, G., Pietrini, P., Robinson, A.,
\& Torricelli, G.\ 2008, \apj, 678, 693



\bibitem[Merloni
\& Heinz(2008)]{2008MNRAS.388.1011M} Merloni, A., \& Heinz, S.\ 2008, \mnras, 388, 1011



\bibitem[Osmer \& Shields(1999)]{1999ASPC..162..235O}
Osmer, P.~S., \& Shields, J.~C.\ 1999,
Quasars and Cosmology, ASPC, 162, 235

\bibitem[Richards et al.(2002)]{2002AJ....124....1R} Richards, G.~T.,
Vanden Berk, D.~E., Reichard, T.~A., Hall, P.~B., Schneider, D.~P.,
SubbaRao, M., Thakar, A.~R., \& York, D.~G.\ 2002, \aj, 124, 1


\bibitem[Richards(2006)]{2006astro.ph..3827R} Richards, G.~T.\ 2006,
arXiv:astro-ph/0603827
(talk presented at the ``AGN Winds in the Caribbean'' Workshop, St. John, USVI;
28 November - 2 December, 2005; http://www.nhn.ou.edu/\~{ }leighly/VImeeting/ )





\bibitem[Shields(2007)]{2007ASPC..373..355S} Shields, J.~C.\ 2007, The
Central Engine of Active Galactic Nuclei, ASPC, 373, 355







\end{thebibliography}
\end{document}